\documentclass[12pt]{JHEP3}
\pdfoutput=1

\usepackage{amsfonts, amsmath, amsthm}
\usepackage{graphicx}
\title{ABJM Dibaryon Spectroscopy}
\author{Jeff Murugan$^{1,2}$\footnote{jeff@nassp.uct.ac.za} ~and Andrea Prinsloo$^{1,2}$\footnote{andy.prinsloo@uct.ac.za}\\

$^{1}$Astrophysics, Cosmology \& Gravity Center and \\
Department of Mathematics and Applied Mathematics, \\
University of Cape Town, \\
Private Bag, Rondebosch, 7700, \\
South Africa.\\

$^{2}$National Institute for Theoretical Physics (NITheP), \\
Private Bag X1, \\
Matieland, 7602, \\
South Africa.}

\abstract{We extend the proposal for a detailed map between wrapped D-branes in Anti-de Sitter space and baryon-like operators in the associated dual conformal field theory provided in \texttt{hep-th/0202150} to the recently formulated $AdS_{4} \times \mathbb{CP}^{3}$/ABJM correspondence. In this example, the role of the dibaryon operator of the 3-dimensional CFT is played by a D4-brane wrapping a $\mathbb{CP}^{2}\subset \mathbb{CP}^{3}$. This topologically stable D-brane in the $AdS_{4}\times \mathbb{CP}^{3}$ is nothing but one-half of the maximal giant graviton on
$\mathbb{CP}^{3}$.}
 \preprint{ACGC-060311}
\newcommand{\p}{\partial}
\begin{document}

\maketitle

\section{Introduction}

D-branes constitute a fundamental piece of the grand puzzle that is string theory. This is true not only because they are honest-to-goodness dynamical degrees of freedom which must be properly accounted for in a consistent formulation of the theory, but also because, by definition, open fundamental strings end on them. Consequently, questions that probe the open-string sector of any string theory will inevitably involve D-branes. This is particularly important in the context of the AdS/CFT correspondence \cite{Maldacena97} where the string physics is encoded in the kinematics and dynamics of 4-dimensional gauge theories (and visa-versa). On the one hand, this has led to enormous advances in the understanding of strongly coupled gauge theories that range from a prediction for a lower bound on the shear viscosity to entropy density of the quark-gluon plasma (see, for example \cite{PSS}) to new understanding of quantum critical points and hydrodynamic behaviour in certain condensed matter systems (see
\cite{Hartnoll} and references therin). On the other, it has provided a concrete and rigorous platform from which to investigate the idea that spacetime, gravity, geometry and topology are all emergent {\it macroscopic} concepts, arising from {\it microscopic} quantum interactions. \\

\noindent
In the context of the $AdS_{5}\times S^{5}$/$\mathcal{N}=4$ SYM (or, as we will often refer to it in this article, the AdS$_{5}$/CFT$_{4}$) correspondence, D-branes in type IIB superstring theory are identified with certain gauge-invariant operators in the dual gauge theory. One class of D-branes in particular has proven to be a very fertile testing ground for many of these ideas: the $\frac{1}{2}$-BPS giant gravitons of \cite{MST00,GMT}. Initially conjectured to be dual to single-trace operators of the form $tr\left(\prod\Phi^{J}\right)$, it was soon realized in \cite{Balasubramanian-et-al} that the correct description of these wrapped D3-branes is in terms of determinant-like operators of the form $det\left(\prod\Phi^{J}\right)$. This is true at least for the so-called {\it maximal giant graviton}, a D3-brane blown up through the Myers effect \cite{Myers} and wrapping
an $S^{3}\subset S^{5}$. For giants not quite at their maximal size, the corresponding dual is identified as a {\it subdeterminant} of the form $\mathcal{O}_{l} = \frac{1}{l!}\epsilon_{i_{1}\cdots i_{l}}
\epsilon^{j_{1}\cdots j_{l}}\Phi^{i_{1}}_{j_{1}}\ldots\Phi^{i_{l}}_{j_{l}}$. Actually, both these operators are just special cases of so-called Schur polynomials (see, for example, \cite{CJR} for the original statement of the giant graviton/Schur polynomial map, \cite{SUN extension,RdMK-et-al} for recent developments in Schur polynomial technology and \cite{JM-RdMK} for an especially readable introduction to some of these ideas), $\chi_{R}(\Phi) = \frac{1}{n!}
\sum_{\sigma\in S_{n}}\chi_{R}(\sigma)\,tr\!\left(\sigma\Phi^{\otimes n}\right)$. A key feature of the Schur operator is that its label $R$ can - via the Schur-Weyl duality - be thought of as a Young diagram with $n$ boxes. Consequently, many detailed dynamical questions in the quantum field theory can be reduced to the combinatorics of Young diagrams. Armed with this formidable technology, by now, an impressive list of successes have emerged. These include (among others):
\begin{itemize}
  \item a description of how the compact topology of the spherical D3-brane is encoded in the gauge
  theory through the Littlewood-Richardson rules for combining Young diagrams
  \cite{RdMK-et-al,Balasubramanian:2004nb},
  \item a demonstration of how the Chan-Paton factors associated with strings ending on multiple
  D-branes arise in a dynamical way in the dual gauge theory \cite{RdMK-et-al,Balasubramanian:2004nb,Carlson-dMK-Lin},
  \item the identification and elucidation of an instability of D-branes to gravitational radiation through
  closed string emission \cite{Berenstein:2006qk} (see also \cite{RdMK-et-al} and
  \cite{Hamilton-Murugan-08} for an application of this idea to braneworld cosmologies),
  \item the realization that the $\frac{1}{2}$-BPS sector of the $\mathcal{N}=4$ SYM is equivalent
  to the dynamics of $N$ free fermions \cite{Berenstein1} that culminated in the LLM
  classification of all $\frac{1}{2}$-BPS type IIB supergravity geometries \cite{Lin:2004nb},
  \item evidence that, like its closed-string counterpart, the open string sector of the type IIB
  superstring is integrable as well \cite{RdMK-et-al,Berenstein:2006qk},
  \item a quantitative description of how geometry emerges in the gauge theory through the study
  of $O(N^{2})$ $\mathcal{R}-$charged operators in the SYM theory \cite{Koch:2008ah} and even,
  \item a proposal for the origins of gravitational thermodynamics \cite{Babel}.
\end{itemize}
Certainly then, the original AdS$_{5}$/CFT$_{4}$ duality appears to furnish an excellent set of tools to answer a variety of questions about the {\it open string} sector of superstring theory. {\it But just how specific are these answers to the AdS$_{5}$/CFT$_{4}$ duality?} Ideally, we'd like the response to be an unequivocal {\it ``not at all"}, but until an actual proof of Maldacena's conjecture \cite{Maldacena97} is found\footnote{See \cite{DJJR} for a promising recent attempt in this direction.} we are probably going to have to settle for a circumstantial {\it ``not as far as we know"}. In this light, it is important that we build up as much circumstantial evidence as possible. This, in turn, necessitates exploring the gauge/gravity duality beyond the type IIB superstring on $AdS_{5}\times S^{5}$ without
loss of the protection of the powerful non-renormalization theorems that accompany the latter.

Fortunately, the AdS$_{4}$/CFT$_{3}$ correspondence recently proposed by Aharony, Bergman, Jafferis and Maldacena (ABJM) \cite{ABJM} does just that. This so-called ABJM duality relates the type IIA superstring on $AdS_{4}\times\mathbb{CP}^{3}$ to (two copies of) an $\mathcal{N}=6$ Super Chern-Simons theory with matter in the bifundamental representation of its $U(N)\times U(N)$ gauge group and level numbers $k$ and $-k$.  Like its better understood AdS$_{5}$/CFT$_{4}$ counterpart, this is also a strong/weak coupling duality with t'Hooft coupling $ \lambda = \frac{N}{k} = \frac{R^{4}}{2\pi^{2}}$, where $R$ denotes the radius of the anti-de Sitter space. Since its discovery in 2008, this particular example has seen an explosion of activity which has unearthed a wealth of new results; and while most of these confirm what we've already learnt in the AdS$_{5}$/CFT$_{4}$ context, there have also been some notable surprises. Among these, of particular interest to us are:
\begin{itemize}
  \item The reported one loop mismatch in the cusp anomalous dimension of string
  states with large angular momentum in $AdS_{4}$ and the associated gauge theory operators
  \cite{Krishnan}.
  Specifically, using the integrability of the scalar sector of the Chern-Simons theory, it can be shown
  that the cusp anomaly is
  \begin{equation}
  f_{CS}(\lambda) = \sqrt{2\lambda} - \frac{3\ln 2}{2\pi} +
  O\left(\frac{1}{\sqrt{\lambda}}\right), \nonumber
  \end{equation}
  while a computation of the energy of a spinning  closed
  string at one loop in the sigma model perturbation energy yields
  \begin{equation}
  f_{string}(\lambda) =
  \sqrt{2\lambda} - \frac{5\ln 2}{2\pi} + O\left(\frac{1}{\sqrt{\lambda}}\right).\nonumber
  \end{equation}
  While several resolutions of this discrepency - including a modification of the regularization
  scheme on the worldsheet \cite{Gromov} as well as one-loop corrections to the
  magnon dispersion relation \cite{McLoughlin} - have been proposed, none are, as yet, fully
  satisfactory and all seem to point to a far more subtle integrability structure in the closed
  string sector than in the $AdS_{5}\times S^{5}$ case \cite{Sundin}.

  \item A rich set of D-brane configurations, some of which have yet to be seen in the
  more familiar type IIB string theory. One such configuration is the {\it toroidal} giant graviton
  of \cite{NT-giants}. Constructed from spinning M2-branes in $AdS_{4}\times S^{7}$ by reducing
  to $AdS_{4}\times \mathbb{CP}^{3}$, these BPS configurations were further analysed in
  \cite{BP} in which the first steps were taken toward identifying the operators in the dual ABJM
  field theory that can be identified with the giant torus. While still incomplete, this very interesting
  proposal nevertheless manages to enumerate all the properties that are necessary for such a giant
  torus operator.
\end{itemize}

Evidently, the AdS$_{4}$/CFT$_{3}$ duality embodied in the ABJM model is {\it not} just one more example of the gauge/gravity correspondence; it is a rich framework within which questions about, for example, the integrability of string theory, the emergence of non-trivial geometry and topology
may be asked or the relation between string theory and M-theory may be unpacked \cite{BLG, MMN}.
{\it We would like to explore the open string sector of the ABJM duality with the aim of understanding its integrability properties and how these translate to M-theory}.  Toward this end, we began our program in \cite{HMPS} by constructing a D2-brane giant graviton\footnote{See also \cite{NT-giants}.} blown up on an $S^{2}\subset AdS_{4}$ and studying in detail its spectrum of small fluctuations. That this type IIA configuration is, modulo dimensionality, identical to the corresponding D3-brane solution in $AdS_{5}\times S^{5}$ in so far as its physical properties go, comes as no surprise\footnote{This is perhaps too weak a statement. As was shown in \cite{HMPS}, the fluctuation spectrum of the D2-brane giant graviton encodes a non-trivial coupling between the transverse fluctuations of the brane and a non-vanishing worldvolume gauge field. This is a new feature of the D2-brane giant.}. Indeed, it is reassuring that the hard-won lessons of the latter carry through to the AdS$_{4}$/CFT$_{3}$ duality. More intriguing, however, is the (Hodge) dual of the D2-brane; the D4-brane giant graviton blown up in the $\mathbb{CP}^{3}$. Not only is this ``${\mathbb{CP}^{3}}$ giant" expected to exhibit quite non-trivial geometry \cite{DG-JM-AP} but, because its field theory dual is essentially known\footnote{This, on the other hand is maybe too strong a statement; it is probably more correct to say that the field theory dual {\it can be guessed at}.}, this means that, building on the considerable technology developed in the AdS$_{5}$/CFT$_{4}$ case \cite{RdMK-et-al}, we can also understand how this geometry is encoded in the ABJM gauge theory. This would constitute a rather non-trivial test of the idea of geometry as an emergent phenomenon. If this sounds too good to be true, it's because it is $\ldots$ at least for now. The problem is that the construction of the giant is not easy. 

The structural similarities of the Klebanov-Witten \cite{Klebanov-Witten} and ABJM field theories, however, led us to believe that insight into this D4-brane giant could be gained by studying a corresponding D3-brane giant graviton on $AdS_{5}\times T^{1,1}$. In \cite{HMP}, we set about constructing just such a D3-brane giant graviton, building on the elegant construction of Mikhailov \cite{Mikhailov} in terms of holomorphic curves on a cone $\mathcal{C} \supset T^{1,1}$ in $\mathbb{C}^{4}$. We were also able to determine the spectrum of small open string fluctuations about this geometry and show perfect matching with the results for the {\it maximal giant} reported in \cite{BHK}. The spectrum exhibited some interesting properties, not least of which is its dependence on the {\it size} of the giant - a signature of the non-trivial geometry of the D-brane worldvolume. Unfortunately, the non-renormalizablilty of the field theory meant that we were unable to complete a comparison of the open string energies with the anomalous dimensions of the dual excited giant graviton operators. This article should be considered the next step in this programme.

In what follows, we extend the study of dibaryons in \cite{BHK} to the ABJM model by constructing the gravitational dual of a baryon-like operator in the {\it renormalizable} 3-dimensional Super Chern-Simons-matter theory. The membrane configuration is a wrapped D4-brane on a topologically non-trivial 4-cycle $\mathbb{CP}^{2}\subset\mathbb{CP}^{3}$ whose field theory dual is a determinant-like operator built out of the bifundamental scalars $A_{i}$ and $B_{i}$. We also construct the spectrum of BPS fluctuations which is then compared with a similar spectrum in \cite{HMP,BHK}. Essentially, we're looking for some indelible imprint of the geometry of the brane worldvolume which we might be able to observe in the dual field theory states. While what we find is not quite so grandiose, it will serve to cement the foundation on which we will build the sequel \cite{DG-JM-AP}.

\section{A summary of dibaryons in Klebanov-Witten theory}

To begin, let's recall some facts about baryon-like operators in Klebanov-Witten theory and their duals; D3-branes wrapping topologically non-trivial cycles in $T^{1,1}$ as identified in \cite{BHK,GK}. Klebanov-Witten  gauge theory \cite{Klebanov-Witten} is an $\mathcal{N}=1$ SYM theory in 4 dimensions, containing two sets of two left-handed chiral superfields $\mathcal{A}_{i}$ and $\mathcal{B}_{i}$ that transform in the $(\mathbf{N},\bar{\mathbf{N}})$ and $(\bar{\mathbf{N}},\mathbf{N})$ representations of the $SU(N)\times SU(N)$ gauge group respectively.  The chiral superfields each carry $R-$charge $\frac{1}{2}$ under the $U(1)_{R}$ symmetry group and consistency requires that an exactly marginal $SU(2)\times SU(2)\times U(1)_{R}$ preserving superpotential of the form.
\begin{equation}
  \mathcal{W}=\tfrac{1}{2}\lambda~\epsilon^{ij}\epsilon^{kl}
  ~\textrm{tr}\left(\mathcal{A}_{i}\mathcal{B}_{k}\mathcal{A}_{j}\mathcal{B}_{l}\right),
\end{equation}
be added. In 4 dimensions, this superpotential is non-renormalizable. Additionally, the scalar components of these chiral multiplets have conformal dimensions $\left[A_{i}\right] = \left[B_{i}\right] = \tfrac{3}{4}$, and carry baryon numbers $1$ and, respectively, $-1$ with respect to the global $U(1)$ symmetry group.

``Dibaryons" in this gauge theory are colour-singlet operators constructed by completely antisymmetrizing with respect to both $SU(N)$'s \cite{BHK,GK}. These come in two types, depending on whether the operator is built out of $A_{i}$'s or $B_{i}$'s. The first,
\begin{eqnarray}
  \mathcal{D}_{1l} = \epsilon_{\alpha_{1}\ldots\alpha_{N}} \hspace{0.1cm} \epsilon^{\beta_  
  {1}\ldots\beta_{N}} \hspace{0.1cm}
  \left[ D_{l}^{k_{1}\ldots k_{N}} \left(A_{k_{1}}\right)^{\alpha_{1}}_{~~\beta_{1}} 
  \ldots \hspace{0.05cm} \left(A_{k_{N}}\right)^{\alpha_{N}}_{~~\beta_{N}} \right],
\end{eqnarray}  
transforms in the $(\mathbf{N+1},\mathbf{1})$ of $SU(2)\times SU(2)$ and carries positive baryon number while the second,
\begin{eqnarray}
  \mathcal{D}_{2l} = \epsilon^{\alpha_{1}\ldots\alpha_{N}} \hspace{0.1cm} \epsilon_{\beta_{1}\ldots\beta_{N}} \hspace{0.1cm}
\left[ D_{l}^{k_{1}\ldots k_{N}}  \left(B_{k_{1}}\right)_{\alpha_{1}}^{~~\beta_{1}} \ldots \hspace{0.05cm} \left(B_{k_{N}}\right)_{\alpha_{N}}^{~~\beta_{N}} \right],
\end{eqnarray}
transforms in the $(\mathbf{1}, \mathbf{N+1})$ with negative baryon number. Both sets of operators have conformal dimension $\Delta = \tfrac{3}{4}$ and $R-$charge $\tfrac{N}{2}$. Much of the physics of the 
dibaryons can be exemplified by state with maximum $J_{3}$ of the first $SU(2)$ which, following \cite{BHK}, we denote as $\det{A_{1}} = \epsilon^{1}\epsilon_{2}\left(A_{1},\ldots,A_{1}\right)\equiv\epsilon^{\alpha_{1}\ldots\alpha_{N}} \epsilon_{\beta_{1}\ldots\beta_{N}} D_{l}^{k_{1}\ldots k_{N}} \prod_{i}^{N} (A_{1})^{\beta_{i}}_{\alpha_{i}}$. Fluctuations about this state produce an excited dibaryon and are formed by replacing one of the $A_{1}$'s with any other chiral field transforming in the same representation of the gauge groups. For instance, making the substitition $A_{1}\rightarrow A_{1}B_{i}A_{j}$ produces a chiral field (up to F-terms) that factorizes into the form $\mathrm{Tr}(B_{i}A_{j})\,\mathrm{det}A_{1}$ {\it i.e.} a gravitational fluctuation on the dibaryon background. On the other hand, the replacement $A_{1}\rightarrow A_{2}B_{i}A_{2}$, results in an operator that does {\it not} factorize into a dibaryon and a single-trace operator. Nevertheless, it is a single-particle state in $AdS_{4}$ and hence a BPS excitation of the D3-brane. More generally, BPS fluctuations of the dibaryon can be studied by replacing one of the $A_{1}$'s by 
$A_{2}\left(B_{i_{1}}A_{2}\right)\ldots \left(B_{i_{n}}A_{2}\right)$. The resulting operator will carry a conformal dimension $\Delta = \tfrac{3}{4}N+\tfrac{3}{2}n$.  We may construct different linear combinations of operators of the above form, but the BPS fluctuation must be the totally symmetric spin $\tfrac{1}{2}n$ state of the global $SU(2)_{B}$ symmetry group. The $SU(2)_{A}$ quantum numbers of the operator, however, remain unknown as far as we are aware.

On the gravity side of the correspondence, Klebanov-Witten theory is dual to type IIB string theory on a 10-dimensional $AdS_{5}\times T^{1,1}$ background.  The compact space $T^{1,1}$ is a $U(1)$ fibration over $S^{2}\times S^{2}$.  Dibaryons are dual to topologically stable D3-branes wrapped on homotopically non-trivial cycles in $T^{1,1}$.  In particular, the determinant operators $\det{A_{1}}$ and $\det{B_{1}}$ correspond to D3-branes wrapped on one of the 2-spheres, $(\theta_{2},\phi_{2})$ or $(\theta_{1},\phi_{1})$ respectively, and the fibre direction $\psi$.

It was shown in \cite{BHK} that small open string excitations of the D3-brane worldvolume in the transverse $T^{1,1}$ directions have eigenfrequencies $\omega^{\pm}_{lms}$, which satisfy
\begin{equation}
\left(\omega^{\pm}_{smp} \pm 2 \right)^{2} = 6l\left(l+1\right) + 3\left(m \mp 1\right)^{2} + 1,
\hspace{1.0cm} \textrm{with}
\hspace{0.5cm} l \equiv s + \textrm{max}\left\{m,|p|\right\}. \hspace{1.0cm}
\end{equation}
Here $m$ and $p$ are either both integer or half-integer (with $m$ taken to be non-negative to remove redundancy), and $s$ is a non-negative integer.  The lowest frequency $s=0$ modes, also satisfying $|p| \leq m$, correspond to eigenfrequencies $\omega^{+} = 3m$, which increase in steps of $\tfrac{3}{2}$ as we vary $m$.  We may associate $m=\tfrac{1}{2}n$ with the spin of the BPS fluctuations. The conformal dimensions then exactly match the lowest eigenfrequencies, when these are added to the energy of the original wrapped D3-brane.

\section{Dibaryons in the ABJM model}

The ABJM model of \cite{ABJM} consists of two copies of a super-Chern-Simons-matter theory in 3 dimensions with level numbers $k$ and $-k$ respectively, $\mathcal{N}=6$ supersymmetry and gauge group $U(N) \times U(N)$.  There are two sets of two chiral superfields $\mathcal{A}_{i}$ and $\mathcal{B}_{i}$ in $\mathcal{N}=2$ superspace \cite{Klose-et-al}, which transform in the $(\mathbf{N},\bar{\mathbf{N}})$ and $(\bar{\mathbf{N}},\mathbf{N})$ bifundamental representations respectively.  The associated scalar fields $A_{i}$ and $B_{i}$ have conformal dimension $[A_{i}]=[B_{i}]=\tfrac{1}{2}$ and may be arranged in the multiplet $Y^{a}=(A_{1},A_{2},B^{\dag}_{1},B^{\dag}_{2})$, with hermitean conjugate $Y^{\dag}_{a} = (A^{\dag}_{1},A^{\dag}_{2},B_{1},B_{2})$. The renormalizable ABJM superpotential takes the form
\begin{equation}
\mathcal{W} = \frac{2\pi}{k} \hspace{0.1cm} \epsilon^{ij} \epsilon^{kl} \hspace{0.1cm}
\textrm{tr}\left(\mathcal{A}_{i}\mathcal{B}_{j}\mathcal{A}_{k}\mathcal{B}_{l}\right),
\end{equation}
and exhibits an explicit $SU(2)_{A}\times SU(2)_{B}$ $\mathcal{R}$-symmetry.  The two $SU(2)$'s act on the doublets
$(A_{1},A_{2})$ and $(B_{1},B_{2})$ respectively.  There is also an additional $SU(2)$ $\mathcal{R}$-symmetry, under which $(A_{1},B^{\dag}_{1})$ and $(A_{2},B^{\dag}_{2})$ transform as doublets, and which enhances the global symmetry group to $SU(4)_{\mathcal{R}}$ under which the $Y^{a}$ transforms in the fundamental representation \cite{Klose-et-al}. 

\subsection{Dibaryon operators}

Dibaryons in the ABJM model may na\"{i}vely be constructed very similarly to those in Klebanov-Witten theory \cite{BHK,GK}. Specifically, the two classes of baryon-like operators take the form
\begin{eqnarray}
&& \mathcal{D}_{1l} = \epsilon_{\alpha_{1}\ldots\alpha_{N}} \hspace{0.1cm} \epsilon^{\beta_{1}\ldots\beta_{N}} \hspace{0.1cm}
\left\{ D_{l,a_{1}\ldots a_{N}} \left(Y^{a_{1}}\right)^{\alpha_{1}}_{~~\beta_{1}} \ldots \hspace{0.05cm} \left(Y^{a_{N}}\right)^{\alpha_{N}}_{~~\beta_{N}} \right\} \label{ABJM-dibaryon1} \\
&& \mathcal{D}_{2l} = \epsilon^{\alpha_{1}\ldots\alpha_{N}} \hspace{0.1cm} \epsilon_{\beta_{1}\ldots\beta_{N}} \hspace{0.1cm}
\left\{ D_{l}^{a_{1}\ldots a_{N}}  \left(Y^{\dag}_{a_{1}}\right)_{\alpha_{1}}^{~~\beta_{1}} \ldots \hspace{0.05cm} \left(Y^{\dag}_{a_{N}}\right)_{\alpha_{N}}^{~~\beta_{N}} \right\}, \label{ABJM-dibaryon2}
\end{eqnarray}
carrying positive and negative baryon number respectively, and have conformal dimension $\Delta = \tfrac{1}{2}N$. For simplicity, in what follows, we shall consider only the determinant operators 
$\det{Y^{1}}=\det{A_{1}}$ and $\det{Y^{\dag}_{3}}=\det{B_{1}}$.

One significant difference between the ABJM and Klebanov-Witten theories is the gauge group, 
$U(N)\times U(N)$, rather than the $SU(N)\times SU(N)$ of Klebanov-Witten. Each $U(N)$ contains an additional local $U(1)$ symmetry, but the current associated with the second $U(1)$ couples only to that of the first $U(1)$ and is hence trivial \cite{ABJM}. The dibaryons
(\ref{ABJM-dibaryon1}) and (\ref{ABJM-dibaryon2}), however, are charged with respect to the extra local $U(1)$ symmetry in ABJM theory and are therefore {\it not gauge invariant} operators. To circumvent this complication, it is possible to attach $N$ Wilson lines - exponentials of integrals over gauge fields - to the dibaryons to make them gauge invariant. Indeed, it was argued in \cite{ABJM} that such operators remain local and that the Wilson lines are unobservable.  This modification should therefore not effect the conformal dimensions of the dibaryons \cite{ABJM,BP}.

\subsection{BPS fluctuations}

Fluctuations about the dibaryon state can be studied along the same lines as in the Klebanov-Witten model. For example, replacing one of the scalar fields in the determinant $\det{A_{1}}$ as follows:
\begin{equation} \label{excitations-detA1}
  A_{1} \rightarrow A_{2}\left(B_{i_{1}}A_{2}\right)\ldots\left(B_{i_{n}}A_{2}\right),
\end{equation}
produces an operator which, in general, will pick up an anomalous dimension. Here, we expect that, for a given $n$, there will exist a linear combination of these operators which remains BPS. This BPS fluctuation will be the totally symmetric spin $\tfrac{1}{2}n$ state of the global $SU(2)_{B} \subset SU(4)_{\mathcal{R}}$ symmetry group. The $SU(2)_{A}$ quantum numbers, however, remain unknown. Similarly, we can replace one of the scalar fields in the determinant operator $\det{B_{1}}$ as follows:
\begin{equation} \label{excitations-detA1}
B_{1} \rightarrow B_{2}\left(A_{i_{1}}B_{2}\right)\ldots\left(A_{i_{n}}B_{2}\right),
\end{equation}
producing another BPS fluctuation that is the totally symmetric spin $\tfrac{1}{2}n$ state of the global $SU(2)_{A} \subset SU(4)_{\mathcal{R}}$ symmetry group. Again, the $SU(2)_{B}$ quantum numbers are unknown but, since the conformal dimensions of these BPS fluctuations of the determinant operators $\det{A_{1}}$ and $\det{B_{1}}$, given by $\Delta = \tfrac{1}{2}N + n$ in both cases, are protected from quantum corrections by supersymmetry, there is some hope that we may compare them to some appropriate quantity on the gravity side of the correspondence. To this end, let's now go to the dual string theory on $AdS_{4}\times \mathbb{CP}^{3}$.

\section{$\mathbb{CP}^{2}$ dibaryons on $AdS_{4}\times\mathbb{CP}^{3}$}

Although by now quite well known, in the interests of self-containment and brevity, we relegate a detailed discussion of the type IIA string on $AdS_{4}\times\mathbb{CP}^{3}$ to appendix \ref{appendix-background}. By way of summary though, let us note that we will make use of the coordinates $(\zeta,\eta, \theta_{i}, \phi_{i})$ - a variation on the parameterization of \cite{NT-giants} - more suited to drawing an analogy with the Einstein space $T^{1,1}$.

\subsection{$\mathbb{CP}^{2}$ dibaryons}

It is possible \cite{GLR} to wrap D4-branes on the two natural $\mathbb{CP}^{2}$ subspaces $(\zeta,\xi,\theta_{2},\phi_{2})$ and $(\zeta,\xi,\theta_{1},\phi_{1})$ of the complex projective space $\mathbb{CP}^{3}$.  Since each of these 4-cycles is non-contractible, these wrapped D4-branes are {\it topologically} stable configurations. By analogy to the Klebanov-Witten case, these D4-branes are expected to be dual to $\det{A_{1}}$ and $\det{B_{1}}$ respectively. The ans\"{a}tze for the two $\mathbb{CP}^{2}$ dibaryons are

\medskip
\begin{tabular}{|p{7.15cm}|p{7.15cm}|}
\hline
\hspace{0.1cm} \underline{\textbf{1st $\mathbb{CP}^{2}$ dibaryon ($\theta_{1}=0$) ansatz}}
& \hspace{0.1cm} \underline{\textbf{2nd $\mathbb{CP}^{2}$ dibaryon ($\theta_{2}=0$) ansatz}} \\
& \\
\hspace{0.1cm} $v_{k} = 0$                        & \hspace{0.1cm} $v_{k} = 0$ \\
\hspace{0.1cm} $\theta \equiv \theta_{1} = 0$     & \hspace{0.1cm} $\theta \equiv \theta_{2} = 0$ \\
\hspace{0.1cm} $\varphi(\sigma^{a}) \equiv \phi_{1}(\sigma^{a})$ unspecified & \hspace{0.1cm} $\varphi(\sigma^{a}) \equiv \phi_{2}(\sigma^{a})$ unspecified \\
& \\
\hspace{0.075cm} \textrm{with worldvolume coordinates} & \hspace{0.075cm} \textrm{with worldvolume coordinates} \\
\hspace{0.1cm} $\sigma^{0} \equiv \tau = t$ & \hspace{0.1cm} $\sigma^{0} \equiv \tau = t$ \\
\hspace{0.1cm} $\sigma^{1} = x \equiv \sin^{2}{\zeta}$         & \hspace{0.1cm} $\sigma^{1} = x \equiv \cos^{2}{\zeta}$ \\
\hspace{0.1cm} $\sigma^{2} = z \equiv \cos^{2}{\tfrac{\theta_{2}}{2}}$ & \hspace{0.1cm} $\sigma^{2} = z \equiv \cos^{2}{\tfrac{\theta_{1}}{2}}$ \\
\hspace{0.1cm} $\sigma^{3} = \xi \equiv \psi + \phi_{1}$ & \hspace{0.1cm} $\sigma^{3} = \xi \equiv \psi + \phi_{2}$ \\
\hspace{0.1cm} $\sigma^{4} = \phi \equiv \phi_{2}$ & \hspace{0.1cm} $\sigma^{4} = \phi \equiv \phi_{1}$ \\
\hline
\end{tabular}
\smallskip

The complex projective space $\mathbb{CP}^{3}$ can now be parameterized in terms of the coordinates $(x,z,\theta,\xi,\phi,\varphi)$, with $(x,z,\xi,\phi)$ on the worldvolume, and $\theta$ and $\varphi$ transverse to the brane. With this, the metric on $\mathbb{CP}^{3}$ can be written as 
\begin{eqnarray}
\nonumber && ds_{\mathbb{CP}^{3}}^{2} = \frac{dx^{2}}{4x\left(1-x\right)}
+ \frac{1}{4} \hspace{0.05cm} x\left(1-x\right)\left[d\xi + \left(2z-1\right)d\phi - \left(1-\cos{\theta}\right)d\varphi\right]^{2} \\
&& \hspace{1.5cm} + \hspace{0.1cm} \left(1-x\right) \left[\frac{dz^{2}}{4z\left(1-z\right)} + z\left(1-z\right)d\phi^{2}\right]
+ \frac{1}{4} \hspace{0.1cm} x\left[d\theta^{2} + \sin^{2}{\theta}d\varphi^{2}\right], \hspace{1.0cm}
\end{eqnarray}
while the 2-form and 6-form field strengths\footnote{Note that the ansatz for the second $\mathbb{CP}^{2}$ dibaryon involves a change in orientation, and the field strength and potential forms pick up an additional minus sign as a result.} take the form
\begin{eqnarray}
\nonumber && F_{2} = - \hspace{0.1cm} \frac{k}{2} \left\{dx \wedge
\left[d\xi + \left(2z-1\right)d\phi - \left(1-\cos{\theta}\right)d\varphi\right] \right. \\
&& \left. \hspace{2.0cm} + \hspace{0.1cm} 2\left(1-x\right) \hspace{0.1cm} dz\wedge d\phi
+ x \sin{\theta} \hspace{0.1cm} d\theta\wedge d\varphi\right\}, \\
\nonumber && \\
&& F_{6} = - \hspace{0.1cm} 3kR^{4} \hspace{0.1cm} x\left(1-x\right)\sin{\theta}
\hspace{0.1cm} d\theta \wedge d\varphi \wedge dx \wedge dz \wedge d\xi \wedge d\phi,
\end{eqnarray}
with associated 1-form and 5-form potentials
\begin{eqnarray}
&& C_{1} = - \hspace{0.1cm} \frac{k}{2} \left\{2\left(1-x\right)\left[d\xi + \left(2z-1\right)d\phi - \left(1-\cos{\theta}\right)d\varphi\right]
- \left(1+\cos{\theta}\right)d\varphi\right\}, \\
&& C_{5} = \hspace{0.1cm} 3kR^{4} \hspace{0.1cm} x\left(1-x\right) \left(1+\cos{\theta}\right)
\hspace{0.1cm} d\varphi \wedge dx \wedge dz \wedge d\xi \wedge d\phi.
\end{eqnarray}

The dynamics of these wrapped D4-branes is encoded in the action $S_{\textrm{D4}} = S_{\textrm{DBI}} + S_{\textrm{WZ}}$, where
\begin{eqnarray}
&& S_{\textrm{DBI}} = -T_{4} \int_{\Sigma}d^{5}\sigma \hspace{0.2cm} e^{-\Phi} \sqrt{-\det\left(\mathcal{P}[g] + \mathcal{F}\right)} \label{action-D4-DBI}, \\
&& S_{\textrm{WZ}} = T_{4} \int_{\Sigma} \hspace{0.1cm} \left\{\mathcal{P}[C_{5}]
+ \mathcal{P}[C_{3}]\wedge \mathcal{F}
+ \frac{1}{2}\hspace{0.1cm} \mathcal{P}[C_{1}]\wedge \mathcal{F} \wedge \mathcal{F} \right\}, \label{action-D4-WZ}
\end{eqnarray}
and $T_{4} \equiv \tfrac{1}{(2\pi)^{4}}$, the D4-brane tension in our units.  Here $\mathcal{F} = 2\pi F$ is proportional to the worldvolume flux $F=dA$ and $\mathcal{P}$ is the pullback to the worldvolume $\Sigma$ of the D4-brane.

N\"{a}ively, one might contemplate completing the ansatz for a $\mathbb{CP}^{2}$ dibaryon by simply turning off the worldvolume flux $F$.  That would be too n\"aive. Indeed, as was noted in \cite{GLR,AHHO}, to cancel the Freed-Witten anomaly that would generically develop in such a wrapped brane configuration, $2\pi F$ must be quantized in half-integer units of the K\"{a}hler form $J$ on $\mathbb{CP}^{3}$. In other words,
\begin{equation}
2\pi F = \left(\tfrac{1}{2} + M\right)\mathcal{P}[J],
\hspace{1.0cm} \textrm{with} \hspace{0.3cm} M \hspace{0.1cm} \epsilon \hspace{0.15cm} \mathbb{Z},
\end{equation}
to cancel the Freed-Witten anomaly \cite{FW}.  It was then argued that, since the relevant quantity appearing in the D4-brane action is actually
$\mathcal{F} = 2\pi F + \mathcal{P}[B]$, turning on a constant NS B-field $B = -\tfrac{1}{2}J$ in the $AdS_{4}\times\mathbb{CP}^{3}$ background would compensate for this effect.  We would then obtain $\mathcal{F} = M\hspace{0.05cm}\mathcal{P}[J]$, with $M=0$ associated with the dibaryon solution.

Although the net result for the action of a $\mathbb{CP}^{2}$ dibaryon
\begin{equation} \label{action-dibaryon}
S_{\textrm{D4}} = -\frac{kR^{4}}{16\pi^{4}} \int_{\Sigma} d^{5}\sigma \hspace{0.1cm} \left(1-x\right)
\left\{1 - 6x \hspace{0.1cm} \dot{\varphi}(\sigma^{a})\right\},
\end{equation}
is the same, the interpretation is quite different.  The field strength $F$ has non-zero spatial components, indicating the existence of magnetic flux on the worldvolume - a result of attaching F1-strings to the D4-brane.  These are dual to the Wilson lines, which it was necessary to attach to the dibaryon operators in ABJM theory to ensure gauge invariance.

The momentum $P_{\varphi}$ conjugate to $\varphi$ and the Hamiltonian $H = P_{\varphi}\hspace{0.1cm}\dot{\varphi} - L$ are given by
\begin{equation}
P_{\varphi} = \frac{kR^{4}}{2\pi^{2}} = N \hspace{1.5cm} \textrm{and} \hspace{1.5cm}
H = \frac{kR^{4}}{4\pi^{2}} = \frac{1}{2}\hspace{0.05cm}N,
\end{equation}
respectively. Here we make use of the correspondence (\ref{thooft}) to express the above quantities in terms of the rank $N$ of the gauge group in the dual ABJM theory. At this point we meet the first check of the proposed dibaryon/determinant correspondence: the energies of either one such $\mathbb{CP}^{2}$ dibaryon precisely matches the conformal dimension $\tfrac{1}{2}N$ of the associated determinant operator.

\subsection{Fluctuation analysis}
That said, it is not entirely surprising that the conformal dimension of the determinant agrees with the energy of the wrapped D4-brane, once all relevant fluxes have been accounted for. A much more non-trivial test would be a matching of the spectrum of small fluctuations about the wrapped brane with the corresponding spectrum of anomalous scaling dimensions of the dual operators. We will provide the spectrum of small fluctuations about a 
$\mathbb{CP}^{2}$ dibaryon here and leave the detailed comparison with the gauge theory for a forthcoming article \cite{DG-JM-AP}.  Both scalar and worldvolume fluctuations will be taken into account, since it is not immediately obvious that the latter decouple.  The ansatz for the scalar fluctuations is
\begin{equation}
v_{k} = \varepsilon\hspace{0.1cm}\delta v_{k}(\sigma^{a}) \hspace{1.25cm} \textrm{and} \hspace{1.25cm}
\theta = \varepsilon\hspace{0.1cm}\delta\theta(\sigma^{a}),
\end{equation}
with $\varphi(\sigma^{a})$ unspecified. More convenient for our purposes, the transverse $\mathbb{CP}^{3}$ coordinates
$y_{1} = \sin{\theta}\cos{\varphi}$ and $y_{2} = \sin{\theta}\sin{\varphi}$, which vanish on the worldvolume of the $\mathbb{CP}^{2}$ dibaryon, are perturbed as follows:
\begin{equation}
y_{i} = \varepsilon \hspace{0.1cm} \delta y_{i}(\sigma^{a}), \hspace{1.2cm} \textrm{with} \hspace{0.5cm}
\delta y_{1} = \delta\theta\cos{\varphi} \hspace{0.5cm} \textrm{and} \hspace{0.5cm}
\delta y_{2} = \delta\theta\sin{\varphi}.
\end{equation}
The worldvolume fluctuation ansatz is $ \mathcal{F} = \varepsilon \hspace{0.1cm} \delta \mathcal{F}(\sigma^{a})$.
Here $\sigma^{a} = (t,x,z,\xi,\phi)$ are the worldvolume coordinates and $\varepsilon$ is a small parameter.

We shall now calculate the D4-brane action, which describes these small fluctuations about the $\mathbb{CP}^{2}$ dibaryon, to quadratic order in the fluctuations.  The DBI action (\ref{action-D4-DBI}) takes the form
\begin{equation}
S_{\textrm{DBI}} = - \frac{1}{16\pi^{4}}\hspace{0.05cm}\frac{k}{2R}
\left\{ \int_{\Sigma} d^{5}\sigma \hspace{0.1cm} \sqrt{-\det{\mathcal{P}[g]}}
+ \frac{1}{2} \hspace{0.05cm} \varepsilon^{2}\int_{\Sigma}  \hspace{0.05cm} \delta \mathcal{F} \wedge * \delta \mathcal{F} \right\},
\end{equation}
with
\begin{eqnarray}
&& \hspace{-0.75cm} \sqrt{-\det{\mathcal{P}[g]}} \approx 2\left(1-x\right) \left\{1 + \frac{1}{2}\hspace{0.05cm}\varepsilon^{2}\hspace{0.05cm}
\sum_{k}\left[\delta v_{k}^{2} - \dot{\delta v_{k}}^{2} + \left(\nabla \delta v_{k}\right)^{2}\right] \right. \\
\nonumber && \hspace{2.6cm} \left. + \hspace{0.08cm} \frac{1}{2}\hspace{0.05cm}\varepsilon^{2} \hspace{0.05cm} x \hspace{0.05cm} \sum_{i}\left[-\dot{\delta y_{i}}^{2} + \left(\nabla \delta y_{i}\right)^{2}\right]
+ \frac{1}{2} \hspace{0.05cm} \varepsilon^{2} \left[\delta y_{2}\left(\p_{\xi}\delta y_{1}\right) - \delta y_{1}\left(\p_{\xi}\delta y_{2}\right)\right]\right\},
\end{eqnarray}
and $*$ the Hodge star operator on the worldvolume of the $\mathbb{CP}^{2}$ dibaryon.  The gradient squared of any function $f(x,z,\xi,\phi)$ on the $\mathbb{CP}^{2}$ subspace can be expressed as
\begin{eqnarray}
&& \hspace{-0.2cm} \left(\nabla f\right)^{2} = x\left(1-x\right)\left(\p_{x}f\right)^{2}
+\frac{1}{x\left(1-x\right)}\left(\p_{\xi}f\right)^{2} \\
\nonumber && \hspace{1.0cm} + \hspace{0.05cm} \frac{1}{\left(1-x\right)}\left\{z\left(1-z\right)\left(\p_{z}f\right)^{2}
+ \frac{1}{4z\left(1-z\right)}\left[\left(2z-1\right)\left(\p_{\xi}f\right) - \left(\p_{\phi}f\right)\right]^{2}\right\}.
\end{eqnarray}

The WZ action takes the form\footnote{Note that, in this WZ action, we need to subtract off similar expressions evaluated at
$\theta = \pi$ (since these terms come from an integral over $\mathbb{CP}^{3}$ of the corresponding field strength forms and the $\theta$ integral runs from $\pi$ to $0$).  This makes no difference to the term involving the 5-form potential, which has been chosen to vanish when $\theta = \pi$, but does result in an additional subtraction from the integral over the 1-form potential.  With this taken into account, the last term results only in a total derivative in the WZ action.}
\begin{equation}
S_{WZ} = \frac{1}{16\pi^{4}} \int_{\Sigma} \hspace{0.1cm} \left\{\mathcal{P}\left[C_{5}\right]
+ \varepsilon R^{2}\hspace{0.1cm}\mathcal{P}\left[C_{3}\right] \wedge \delta \mathcal{F}
+ \frac{1}{2} \hspace{0.05cm} \varepsilon^{2} R^{4}\hspace{0.1cm}\mathcal{P}\left[C_{1}\right] \wedge \delta \mathcal{F} \wedge \delta \mathcal{F}\right\},
\end{equation}
where the potentials, pulled back to the worldvolume of the D4-brane, are given by
\begin{eqnarray}
\nonumber && \!\!\! \mathcal{P}\left[C_{5}\right] = 6kR^{4} \hspace{0.05cm} x\left(1-x\right)
\left[\dot{\varphi} - \frac{1}{4} \hspace{0.05cm} \varepsilon^{2}
\left(\delta y_{2} \dot{\delta y_{1}} - \delta y_{1} \dot{\delta y_{2}}\right)
+ O(\varepsilon^{4})\right] dt \wedge dx \wedge dz \wedge d\xi \wedge d\phi \\
\nonumber && \!\!\! \mathcal{P}\left[C_{3}\right] = O(\varepsilon^{3}) \\
&& \!\!\! \mathcal{P}\left[C_{1}\right] = -\frac{1}{2} \hspace{0.05cm} k\left(1-x\right)\left[d\xi + \left(2z-1\right)d\phi\right]
+ k \left(\p_{a}\varphi\right) d\sigma^{a} + O(\varepsilon^{2}).
\end{eqnarray}

Combining the DBI and WZ terms and expanding in $\epsilon$, the D4-brane action can be approximated by
$S_{\textrm{D4}} = S_{0} + \varepsilon^{2}S_{2} + \ldots$,
with $S_{0}$ the original action (\ref{action-dibaryon}) for the $\mathbb{CP}^{2}$ dibaryon and
\begin{eqnarray}
\nonumber && \!\!\! S_{2} = -\frac{kR^{4}}{32\pi^{4}}\int_{\Sigma} d^{5}\sigma \hspace{0.1cm} \left(1-x\right)
\left\{\sum_{k}\left[\delta v_{k}^{2} - \dot{\delta v_{k}}^{2} + \left(\nabla\delta v_{k}\right)^{2}\right] + x \hspace{0.05cm}
\sum_{i}\left[- \dot{\delta y_{i}}^{2} + \left(\nabla\delta y_{i}\right)^{2}\right] \right. \\
\nonumber && \hspace{5.55cm} \left. -3x\left(\delta y_{2} \dot{\delta y_{1}} - \delta y_{1} \dot{\delta y_{2}}\right)
+ \left[\delta y_{2}\left(\p_{\xi}\delta y_{1}\right) - \delta y_{1}\left(\p_{\xi}\delta y_{2}\right)\right]\right\} \\
&& \hspace{1.155cm} - \frac{1}{32\pi^{4}}\hspace{0.05cm}\frac{k}{2R} \int_{\Sigma} \hspace{0.1cm} \delta \mathcal{F} \wedge * \delta \mathcal{F},
\end{eqnarray}
encoding the quadratic order corrections.  Notice that the worldvolume fluctuations decouple. We shall therefore henceforth confine our attention to the scalar fluctuations\footnote{Fermions on the D-brane can be related to the scalar fluctuations by supersymmetry.}. The equations of motion for $\delta v_{k}$ and $\delta y_{\pm}$, which are the combinations $\delta y_{\pm} \equiv \delta y_{1} \pm i\delta y_{2}$ of the transverse $\mathbb{CP}^{3}$ fluctuations, are
\begin{eqnarray}
&& \ddot{\delta v_{k}} - \nabla^{2} \delta v_{k} + \delta v_{k} = 0, \label{eom-ads} \\
&& \ddot{\delta y_{\pm}} \mp 3i\dot{\delta y_{\pm}} - \nabla^{2} \delta y_{\pm}
- \left(1-x\right) \left(\p_{x}\delta y_{\pm}\right) \pm \frac{i}{x}\left(\p_{\xi}\delta y_{\pm}\right) = 0, \label{eom-cp3}
\end{eqnarray}
with the $\mathbb{CP}^{2}$ Laplacian given by (\ref{laplacian}).

If we expand the $AdS_{4}$ fluctuations in terms of the complete set of chiral primaries $\chi_{l}$ defined in (\ref{chiral-primary}), solutions to the equations of motion (\ref{eom-ads}) take the form
\begin{equation}
\delta v_{k} = \sum_{l} e^{-i\omega^{k}_{l}\hspace{0.05cm}t} \hspace{0.15cm} \chi_{l}\left(z^{A},\bar{z}_{B}\right),
\end{equation}
with frequencies
\begin{equation}
\left(\omega^{k}_{l}\right)^{2} = l\left(l+2\right) + 1,
\end{equation}
where $l$ is a non-negative integer.

It turns out, however, that the chiral primaries are not a suitable set of functions over which to expand the transverse $\mathbb{CP}^{3}$ fluctuations.  We rather make use of both sets of eigenfunctions
$\Phi^{\pm}_{smp}(z,x,\xi,\phi)$, written down explicitly in (\ref{dibaryon-mod-eigenfunctions}), of modified operators $\mathcal{O}_{\pm}$. The transverse fluctuations are then given by
\begin{equation} \label{fluctuations-y+-}
\delta y_{\pm} = \sum_{s,m,p} e^{-i\omega^{\pm}_{smp}\hspace{0.05cm}t}~\Phi^{\pm}_{smp}\left(x,z,\xi,\phi\right)
+ \sum_{s,m,p} e^{+i\omega^{\mp}_{smp}\hspace{0.05cm}t}~\Phi^{\mp}_{smp}\left(x,z,-\xi,-\phi\right),
\end{equation}
which solve the equations of motion (\ref{eom-cp3}), if the frequencies satisfy
\begin{equation}
\omega^{+}_{smp}\left(\omega^{+}_{smp}+3\right) = l\left(l+3\right) \hspace{0.75cm} \textrm{and} \hspace{0.75cm}
\omega^{-}_{smp}\left(\omega^{-}_{smp}-3\right) = l\left(l+3\right) + 2,
\end{equation}
with $l = s + 2m$. Here $s$ is an non-negative integer, and $m$ and $p$ are either both integer or half-integer, with $m \geq |p|$.  Also $m$ has been chosen to be non-negative to remove redundancy in (\ref{fluctuations-y+-}).  The lowest frequency mode with $s=0$ has $\omega^{+} = 2m$, simply increasing in integer steps as we change $m$. We may again associate $m=\tfrac{1}{2}n$ with the spin of the BPS fluctuations.  The conformal dimensions $\Delta = \tfrac{1}{2}N + n$ of these fluctuations then exactly match the lowest eigenfrequencies, when these are added to the energy $\tfrac{1}{2}N$ of the original $\mathbb{CP}^{3}$ dibaryon.

\section{Discussion}

At this junction, there are several points to be made and just as many questions raised: 
\begin{itemize}
\item
To begin, it seems clear to us from this analysis that the
dibaryons in the 3-dimensional ABJM theory should correspond to D4-branes wrapping a non-contractible $\mathbb{CP}^{2}\subset \mathbb{CP}^{3}$. Actually, following the computations in \cite{BHK}, it is easy to see that the moduli space of this wrapped D4-brane is itself a $\mathbb{CP}^{3}$. Since the brane couples to the $F_{6}$ flux, a particle on the moduli space is charged under a magnetic field with $N$ units of flux and its wavefunction must be a section of the $N^{\textrm{th}}$ symmetric product of the charged line bundle over $\mathbb{CP}^{3}$. There are
$\left(^{N+3}_{\hspace{0.2cm}N}\right)$ such sections, matching nicely the dimension of the $N^{\textrm{th}}$ symmetric representation of $SU(4)$ {\it i.e.} the dibaryon operator.
\item
In particular, operators of the form $Y^{N}$, like $\mathcal{D}_{l1}$ and $\mathcal{D}_{l2}$ in (\ref{ABJM-dibaryon1}) and (\ref{ABJM-dibaryon2}) respectively, carry $U(1)$ charge $N$ and each correspond to a {\it single} wrapped D4-brane. We have, by studying the eigenvalue problem on $\mathbb{CP}^{2}$, extracted the spectrum of BPS fluctuations about the wrapped D4-brane and showed that, at least qualitatively, there is excellent agreement between the lowest eigenfrequencies and the conformal dimensions of BPS excitations about the dibaryon state in the CFT. A full quantitative matching of the spectra may still require the determining of the $SU(2)_{B}$ quantum numbers but the 3-dimensional ABJM model, unlike the 4-dimensional Klebanov-Strassler theory, is at least renormalizable. 

\item
The lift of this D4-brane to M-theory should be a twisted M5-brane wrapped on a single $\mathbb{Z}_{k}$ cycle of $S^{1}/\mathbb{Z}_{k} \times \mathbb{CP}^{2}$. Clearly, up to $k-1$ such branes may be wrapped before returning to the trivial cycle. It would be interesting to see how this $\mathbb{Z}_{k}$ charge can be measured in the dual gauge theory.

\item
This instability of $k$ D4-branes can be seen in the string theory by the fact that there should be an NS5-brane instanton that turns $k$ D4-branes into $N$ D0-branes. In the gauge theory, this translates into the statement that we need to consider representations labelled by Young diagrams with $N$ rows and $k$ columns (where each column is a D4-brane). Indeed, as was suggested in \cite{ABJM} already,
$k$ scalar fields in the completely symmetric representation (a single row of boxes in the Young diagram)
give a single D0-brane, with an t' Hooft operator $\displaystyle W_{k(\mu_{i},\mu_{j})} = \mathcal{P}\exp\left(i\int_{0}^{\infty}A_{1}^{k\mu_{i}} + A_{2}^{k\mu_{j}}\right)$ attached to ensure gauge invariance. Naively, it would seem that two independent D0-branes should be dual to an operator in the 
$2k\mu_{1}$ representation rather than the $k\mu_{2}$. It remains to be seen how these representations fit together.

\item
On the other hand, dibaryons are not the only gauge-invariant operators that may be constructed from the scalar sector of the ABJM model. Another class of operator of special interest are multi-trace operators built from a composite scalar field of the form $\displaystyle  (Y^{a}Y^{\dag}_{b})^{\alpha}_{\beta}$ which transforms in the $(1,0,1)$ of $SU(4)$. In particular, $U(1)$ neutral Schur polynomials $\chi_{R}(Y^{a}Y^{\dag}_{b})$, where $R$ is a large representation of $SU(N)$, which transform in the $(J,0,J)$ of $SU(4)$, should correspond to moving, non-topological D4-branes in $\mathbb{CP}^{3}$, {\it i.e. giant gravitons}. In fact, as noted in \cite{BT}, the maximal such giant graviton should correspond to the operator combination dibaryon/anti-dibaryon, a reflection of the factorization $\det(A_{1}B_{1}) = \det{A_{1}}\det{B_{1}}$. Exactly how this factorization happens is an interesting problem in its own right and one that we return to in the sequel to this article \cite{DG-JM-AP}. 

\end{itemize}
\section{Acknowledgements}

We would like to thank Robert de Mello Koch for initially suggesting this problem to us, Per Sundin for useful discussions and especially Alex Hamilton for collaboration on the early stages of the work. This work is
based upon research supported by the National Research Foundation (NRF) of South Africa's Thuthuka and Key International Scientific Collaboration programs. AP is supported by an NRF Innovation Postdoctoral Fellowship. Any opinions, findings and conclusions or recommendations expressed in this material are those of the authors and therefore the NRF do not accept any liability with regard thereto.

\appendix

\section{Type IIA string theory on $AdS_{4}\times \mathbb{CP}^{3}$} \label{appendix-background}

Type IIA string theory on the background $AdS_{4}\times \mathbb{CP}^{3}$ has the metric
\begin{equation}
R^{-2}ds^{2} = ds_{AdS_{4}}^{2} + 4ds_{\mathbb{CP}^{3}}^{2},
\end{equation}
while the dilaton $\Phi$ satisfies $e^{2\Phi} = \tfrac{4R^{2}}{k^{2}}$ and the field strength forms are given by
\begin{eqnarray}
&& F_{2} = 2kJ \\
&& F_{4} = -\tfrac{3}{2}kR^{2} ~ \textrm{vol}\left(AdS_{4}\right) \\
&& F_{6} = *F_{4} = \tfrac{3}{2}(64)kR^{4} ~ \textrm{vol}\left(\mathbb{CP}^{3}\right) \\
&& F_{8} = *F_{2},
\end{eqnarray}
where $J$ is the K\"{a}hler form on $\mathbb{CP}^{3}$.  The length scale $R$ of the spacetime is related to the t'Hooft coupling $\lambda = \tfrac{N}{k}$ of ABJM theory as follows\footnote{Here we make use of units in which $\alpha' = 1$.}:
\begin{equation} \label{thooft}
R^{2} = \pi\sqrt{2\lambda},
\end{equation}
and the constant integer $k$, which results from the compactification of M-theory on $AdS_{4}\times S^{7}/\mathbb{Z}_{k}$ to type IIA string theory on $AdS_{4}\times \mathbb{CP}^{3}$, is the SCS-matter theory level number.

\subsection{Anti-de Sitter ($AdS_{4}$) spacetime}

The $AdS_{4}$ metric is given by
\begin{equation}
ds_{AdS_{4}}^{2} = -\left(1+r^{2}\right)dt^{2} + \frac{dr^{2}}{\left(1 + r^{2}\right)} + r^{2}\left(d\theta^{2} + \sin^{2}{\theta} d\varphi^{2}\right).
\end{equation}
The 4-form field strength of the $AdS_{4}\times\mathbb{CP}^{3}$ background is then
\begin{equation}
F_{4} = -\frac{3}{2} \hspace{0.05cm} kR^{2}r^{2}\sin{\theta} \hspace{0.1cm} dt \wedge dr \wedge d\theta \wedge d\varphi,
\hspace{0.5cm} \textrm{with} \hspace{0.3cm}
C_{3} = \frac{1}{2} \hspace{0.05cm} kR^{2}r^{3}\sin{\theta} \hspace{0.1cm} dt \wedge d\theta \wedge d\varphi
\end{equation}
the associated 3-form potential.

Let us now define an alternative set of cartesian coordinates $v_{k}$, which parameterize the anti-de Sitter spacetime, as follows:
\begin{equation}
v_{1} = r\cos{\theta}, \hspace{1.0cm}
v_{2} = r\sin{\theta}\cos{\varphi} \hspace{1.0cm} \textrm{and} \hspace{1.0cm}
v_{3} = r\sin{\theta}\sin{\varphi},
\end{equation}
in terms of which the $AdS_{4}$ metric can be written as
\begin{equation}
ds_{AdS_{4}}^{2} = -\left(1 + \sum_{k}v_{k}^{2}\right)dt^{2}
+ \sum_{i,j}\left(\delta_{ij} - \frac{v_{i}v_{j}}{\left(1 + \sum_{k}v_{k}^{2}\right)}\right)dv_{i}dv_{j}.
\end{equation}
The 4-form field strength becomes
\begin{equation}
F_{4} = -\frac{3}{2} \hspace{0.05cm} kR^{2} \hspace{0.1cm} dt \wedge dv_{1} \wedge dv_{2} \wedge dv_{3},
\end{equation}
which corresponds to the 3-form potential
\begin{equation}
C_{3} = \frac{1}{2} \hspace{0.05cm} kR^{2} \hspace{0.1cm} dt \wedge
\left( v_{1} dv_{2} \wedge dv_{3} + v_{2} dv_{3} \wedge dv_{1} + v_{3} dv_{1} \wedge dv_{2}\right).
\end{equation}

\subsection{Complex projective space ($\mathbb{CP}^{3}$)}\label{section-cp3}

The complex projective space $\mathbb{CP}^{3}$ is described by the homogenous coordinates $z^{A} $ of $\mathbb{C}^{4}$, which are identified up to an overall magnitude (so confined to $S^{7}$) and phase.  These homogenous coordinates can be parametrized\footnote{We make use of a variation of the parametrization of \cite{NT-giants} with an analogy to $T^{1,1}$ in mind.} as follows:
\begin{eqnarray} \label{homogenous-cp3}
\nonumber && z^{1} = \cos{\zeta}\sin{\tfrac{\theta_{1}}{2}} ~ e^{i\left(y + \frac{1}{4}\psi - \frac{1}{2}\varphi_{1}\right)} \hspace{1.0cm}
z^{2} = \cos{\zeta}\cos{\tfrac{\theta_{1}}{2}} ~ e^{i\left(y + \frac{1}{4}\psi + \frac{1}{2}\phi_{1}\right)} \\
&& z^{3} = \sin{\zeta}\sin{\tfrac{\theta_{2}}{2}} ~ e^{i\left(y - \frac{1}{4}\psi + \frac{1}{2}\phi_{2}\right)} \hspace{1.05cm}
z^{4} = \sin{\zeta}\cos{\tfrac{\theta_{2}}{2}} ~ e^{i\left(y - \frac{1}{4}\psi - \frac{1}{2}\phi_{2}\right)},
\end{eqnarray}
with $\zeta \in [0,\tfrac{\pi}{2}]$, $\theta_{i} \in [0,\pi]$, $\psi \in [0,4\pi]$ and $\phi_{i} \in [0,2\pi]$.  Here
$y \in [0,2\pi]$ is the total phase on which the inhomogenous coordinates $\tfrac{z^{i}}{z^{4}}$ of $\mathbb{CP}^{3}$ do not depend.

In these coordinates $\zeta$, $\theta_{i}$, $\psi$ and $\phi_{i}$, the Fubini-Study metric of $\mathbb{CP}^{3}$ is given by
\begin{eqnarray}
\nonumber & ds_{\mathbb{CP}^{3}}^{2} = & d\zeta^{2}
+ \frac{1}{4}\hspace{0.05cm}\cos^{2}{\zeta}\sin^{2}{\zeta}
\left[d\psi + \cos{\theta_{1}}d\phi_{1} + \cos{\theta_{2}}d\phi_{2}\right]^{2} \\
&& + \frac{1}{4}\hspace{0.05cm}\cos^{2}{\zeta}\left(d\theta_{1}^{2} + \sin^{2}{\theta_{1}}d\phi_{1}^{2}\right)
+ \frac{1}{4}\hspace{0.05cm}\sin^{2}{\zeta}\left(d\theta_{2}^{2} + \sin^{2}{\theta_{2}}d\phi_{2}^{2}\right),
\end{eqnarray}
whereas the 2-form and 6-form field strengths of the $AdS_{4}\times \mathbb{CP}^{3}$ background can be written as
\begin{eqnarray}
\nonumber && F_{2} = -\frac{k}{2} \left\{2\sin{\zeta}\cos{\zeta} \hspace{0.1cm} d\zeta
\wedge \left[d\psi + \cos{\theta_{1}}d\phi_{1} + \cos{\theta_{2}}d\phi_{2}\right] \right. \\
&& \left. \hspace{2.0cm} + \cos^{2}{\zeta}\sin{\theta_{1}} \hspace{0.1cm} d\theta_{1} \wedge d\phi_{1}
- \sin^{2}{\zeta}\sin{\theta_{2}} \hspace{0.1cm} d\theta_{2} \wedge d\phi_{2} \right\} \\
\nonumber && \\
&& F_{6} = 3kR^{4} \cos^{3}{\zeta}\sin^{3}{\zeta}\sin{\theta_{1}}\sin{\theta_{2}}
\hspace{0.1cm} d\zeta \wedge d\theta_{1} \wedge d\theta_{2} \wedge d\psi \wedge d\phi_{1} \wedge d\phi_{2}.
\end{eqnarray}

\section{Eigenvalue problems on $\mathbb{CP}^{2}$} \label{appendix-cp2}

\subsection{Embedding $\mathbb{CP}^{2} \subset \mathbb{CP}^{3}$} \label{appendix-cp2-embedding}

The dibaryons are wrapped on a $\mathbb{CP}^{2}$ subspace, which is parameterized by the coordinates $x,\hspace{0.05cm} z \in [0,1]$, $\xi \in [0,4\pi]$ and $\phi \in [0,2\pi]$. The homogenous coordinates of $\mathbb{CP}^{2}$ take the form
\begin{equation}
z^{1} = x \hspace{0.15cm} e^{\frac{1}{2}i\xi}, \hspace{0.7cm}
z^{2} = \sqrt{\left(1-x\right)\left(1-z\right)} \hspace{0.2cm} e^{\frac{1}{2}i\phi}
\hspace{0.5cm} \textrm{and} \hspace{0.5cm}
z^{3} = \sqrt{\left(1-x\right)z} \hspace{0.2cm} e^{-\frac{1}{2}i\phi},
\end{equation}
which can be obtained (up to an overall phase and an interchange of the $z^{A}$'s) from those of $\mathbb{CP}^{3}$, given in
(\ref{homogenous-cp3}), making use of our ansatz for either $\mathbb{CP}^{2}$ dibaryon. The Fubini-Study metric of $\mathbb{CP}^{2}$ is
\begin{equation}
ds_{\mathbb{CP}^{2}}^{2} = \frac{dx^{2}}{4x\left(1-x\right)}
+ \frac{1}{4} \hspace{0.05cm} x\left(1-x\right)\left[d\xi + \left(2z-1\right)d\phi\right]^{2}
+ \hspace{0.1cm} \left(1-x\right) \left[\frac{dz^{2}}{4z\left(1-z\right)} + z\left(1-z\right)d\phi^{2}\right].
\end{equation}

\subsection{Laplacian and chiral primaries} \label{appendix-cp2-chiral-primaries}

Following \cite{MZ-SCS}, we shall write down the Laplacian and chiral primaries in terms of the homogenous coordinates of $\mathbb{CP}^{2}$.  Let us first define the Laplace-Beltrami operator
\begin{equation}
L^{A}_{~B} \equiv z^{A}~\frac{\p}{\p z^{B}} - \bar{z}_{B}~\frac{\p}{\p \bar{z}_{A}},
\end{equation}
in terms of which the Laplacian can be written as
\begin{eqnarray}
& \!\! \nabla^{2} & \equiv -\frac{1}{2}\sum_{A,B} L^{A}_{~B}~L^{B}_{~A} \\
\nonumber && = -\frac{1}{2}\sum_{A,B} \left\{ z^{A}z^{B} \frac{\p}{\p z^{A}}\frac{\p}{\p z^{B}} + \bar{z}_{A}\bar{z}_{B} \frac{\p}{\p \bar{z}_{A}}\frac{\p}{\p \bar{z}_{B}} \right\} + \sum_{A}\frac{\p}{\p z^{A}}\frac{\p}{\p \bar{z}_{A}}
- \frac{3}{2}\sum_{A}\left\{z^{A}\frac{\p}{\p z^{A}} + \bar{z}_{A}\frac{\p}{\p \bar{z}_{A}}\right\}.
\end{eqnarray}
Any function on $\mathbb{CP}^{2}$ can be expanded in terms of the chiral primaries
\begin{equation} \label{chiral-primary}
\chi_{l}\left(z^{A},\bar{z}_{B}\right) = \sum_{A_{i},B_{i}} \chi_{A_{1}\ldots A_{l}}^{B_{1}\ldots B_{l}} ~ z^{A_{1}} \ldots z^{A_{l}} ~ \bar{z}_{B_{1}} \ldots \bar{z}_{B_{l}},
\end{equation}
with $\chi_{A_{1}\ldots A_{l}}^{B_{1}\ldots B_{l}}$ symmetric (under interchange of any two $A_{i}$ or $B_{i}$) and traceless. These are eigenfunctions of the Laplacian on $\mathbb{CP}^{2}$:
\begin{equation} \label{laplacian-eigenvalues1}
\nabla^{2} \chi_{l} = -l\left(l+2\right)\chi_{l},
\end{equation}
where the eigenvalues are dependent only on the length $l$.

\subsection{Stationary eigenvalue problems} \label{appendix-cp2-evps}

The standard stationary eigenvalue problem $\nabla^{2}\Phi = -E\Phi$ on the complex projective space $\mathbb{CP}^{2}$ can be solved using the chiral primaries
(\ref{chiral-primary}) with eigenvalues (\ref{laplacian-eigenvalues1}).  However, we shall rather describe the $\mathbb{CP}^{2}$ subspace using the coordinates $(x,z,\xi,\phi)$, in terms of which the Laplacian can be written as
\begin{eqnarray} \label{laplacian}
\nonumber && \nabla^{2} \equiv \p_{x}\left[x\left(1-x\right)\p_{x}\right] - x\hspace{0.1cm}\p_{x}
+ \frac{1}{x\left(1-x\right)} \hspace{0.1cm} \p_{\xi}^{2} \\
&& \hspace{1.08cm} + \hspace{0.08cm}\frac{1}{\left(1-x\right)}\left\{\p_{z}\left[z\left(1-z\right)\p_{z}\right]
+ \frac{1}{4z\left(1-z\right)}\left[\left(2z-1\right)\p_{\xi}-\p_{\phi}\right]^{2}\right\}, \hspace{1.0cm}
\end{eqnarray}
and look for separable solutions.  This method is then applied to the modified eigenvalue problem $\mathcal{O}_{\pm}\Phi = -E\Phi$, where we define
\begin{equation}
\mathcal{O}_{\pm} \equiv \nabla^{2} + \left(1-x\right)\p_{x} \mp \frac{i}{x} \hspace{0.05cm} \p_{\xi},
\end{equation}
which is associated with transverse $\mathbb{CP}^{3}$ fluctuations.  The solutions involve hypergeometric functions, which are similar in nature to those
discussed in \cite{BHK,Gubser}.

\bigskip
\emph{\textbf{Standard eigenvalue problem}}
\bigskip

Let us consider separable solutions to the standard eigenvalue problem $\nabla^{2}\Phi = -E\Phi$ of the form
\begin{equation} \label{separable}
\Phi(x,z,\xi,\phi) = f(z) \hspace{0.05cm} g(z) \hspace{0.1cm} e^{im\xi} \hspace{0.1cm} e^{ip\phi},
\end{equation}
with\footnote{Functions in $\mathbb{CP}^{2}$ are built out of equal numbers of $z$'s and $\bar{z}$'s - an excess of $z^{1}$'s must be accounted for by no more $\bar{z}_{2}$ or $\bar{z}_{3}$'s (and similarly for an excess of $\bar{z}_{1}$'s).} $\left|m\right| \geq \left|p\right|$.  Here $m$ and $p$ must be either both integer or both half-integer\footnote{This is obtained by reverting to the original coordinates $\psi \in [0,4\pi]$ and $\phi \in [0,2\pi]$ with $\xi = \psi - \phi$:
\begin{equation}
e^{im\xi}e^{-ip\phi} = e^{im\psi}e^{i(p-m)\phi},
\end{equation}
with $m$ and integer or half-integer, and the difference $p-m$ an integer.}.  Note that, if $\Phi(x,z,\xi,\phi)$ is an eigenfunction with eigenvalue $E$, then $\Phi(x,z,-\xi,-\phi)$ is also an eigenfunction with the same eigenvalue.  Thus it is sufficient to consider positive $m$ only, since the negative $m$ solutions can be obtained by a reflection in $\xi$ and $\phi$.

We must now solve two related eigenvalue problems associated with $g(z)$ and $f(x)$, the first of which is given by
\begin{equation}
\p_{z}\left[z\left(1-z\right)\left(\p_{z}g\right)\right]
- \left\{\frac{\left(m+p\right)^{2}}{4} \hspace{0.05cm} \frac{\left(1-z\right)}{z}
+ \frac{\left(m-p\right)^{2}}{4}\frac{z}{\left(1-z\right)} + \frac{\left(p^{2}-m^{2}\right)}{2}
- \lambda\right\}g = 0,
\end{equation}
with $\lambda$ some constant eigenvalue. Setting
\begin{equation}
g(z) = z^{\frac{1}{2}|m+p|}\left(1-z\right)^{\frac{1}{2}|m-p|} \hspace{0.05cm} h_{1}(z),
\end{equation}
we obtain the hypergeometric differential equation
\begin{eqnarray}
\nonumber && z\left(1-z\right)\p_{z}^{2}h_{1} + \left[\left(|m+p|+1\right) - \left(|m+p| + |m-p|+2\right)z\right]\p_{z}h_{1} \\
&& - \left\{\tfrac{1}{2}|m^{2}-p^{2}| + \tfrac{1}{2}|m+p| + \tfrac{1}{2}|m-p| + \tfrac{1}{2}\left(m^{2}-p^{2}\right)
- \lambda\right\}h_{1} = 0. \hspace{1.5cm}
\end{eqnarray}
Solutions take the form of hypergeometric functions $h_{1}(z) = F(a_{1},b_{1},c_{1};z)$, which are dependent on the parameters
\begin{equation}
a_{1}, \hspace{0.05cm} b_{1} \equiv \tfrac{1}{2}|m+p| + \tfrac{1}{2}|m-p| + \tfrac{1}{2} \pm \sqrt{\lambda + m^{2} + \tfrac{1}{4}}
\hspace{0.5cm} \textrm{and} \hspace{0.5cm}
c_{1} \equiv |m+p| + 1,
\end{equation}
where $a_{1}$ and $b_{1}$ are associated with different signs in the $\pm$. For regularity at $z=1$, either $a_{1}$ or $b_{1}$ should be a negative integer.  Hence
\begin{equation}
\tfrac{1}{2}|m+p| + \tfrac{1}{2}|m-p| + \tfrac{1}{2} - \sqrt{\lambda + m^{2} + \tfrac{1}{4}} = -s_{1},
\hspace{0.75cm}
\textrm{with} \hspace{0.4cm} s_{1}  \in \left\{0, \hspace{0.05cm} 1, \hspace{0.05cm}2, \hspace{0.05cm} \ldots\right\},
\end{equation}
so that $\lambda = q\left(q+1\right) - m^{2}$, where we define $q \equiv s_{1} + m$.

The second eigenvalue problem then becomes
\begin{equation}
\p_{x}\left[x\left(1-x\right)\left(\p_{x}f\right)\right] - x\left(\p_{x}f\right)
- \left\{\frac{m^{2}\left(1-x\right)}{x} + \frac{q\left(q+1\right)x}{\left(1-x\right)}
+ m^{2} + q\left(q+1\right) - E\right\}f = 0.
\end{equation}
We shall now take
\begin{equation}
f(x) = x^{m}\left(1-x\right)^{q} \hspace{0.05cm} h_{2}(x),
\end{equation}
where $h_{2}(x)$ satisfies the hypergeometric differential equation
\begin{eqnarray}
\nonumber && x\left(1-x\right)\p_{x}^{2}h_{2} + \left[\left(2m+1\right) - \left(2m + 2q + 3\right)x\right]\p_{x}h_{2} \\
&& - \left\{2\left(mq + m\right) + q\left(q+1\right) + m^{2} - E \right\}h_{2} = 0.
\end{eqnarray}
The solutions $h_{2}(x) = F(a_{2},b_{2},c_{2};x)$ are hypergeometric functions dependent on the parameters
\begin{equation}
a_{2}, \hspace{0.05cm} b_{2} = m + q + 1 \pm \sqrt{E + 1}
\hspace{1.0cm} \textrm{and} \hspace{1.0cm}
c_{2} = 2m + 1. \hspace{1.0cm}
\end{equation}
Here we obtain that $E = l\left(l+2\right)$, where $l \equiv s_{2} + q + m$, with $s_{2} \in \left\{0, \hspace{0.05cm} 1, \hspace{0.05cm}2, \hspace{0.05cm} \ldots\right\}$, for regularity at $x=1$.

Hence the eigenfunctions of the $\mathbb{CP}^{2}$ Laplacian are
\begin{equation}
\Phi_{smp}\left(x,z,\xi,\phi\right) = z^{\frac{1}{2}|m+p|}\left(1-z\right)^{\frac{1}{2}|m-p|} x^{m}\left(1-x\right)^{s_{1}+m}
F^{\hspace{0.05cm} z}_{s_{1}mp}(z) \hspace{0.1cm} F^{\hspace{0.05cm} x}_{s_{2}mp}(x) \hspace{0.1cm} e^{im\xi} \hspace{0.1cm} e^{ip\phi},
\end{equation}
which correspond to the eigenvalues
\begin{equation}
E = l\left(l+2\right), \hspace{1.25cm} \textrm{with} \hspace{0.5cm} l = s + 2m,
\end{equation}
where $s = s_{1} + s_{2}$ is a non-negative integer.  The hypergeometric functions defined by
$F^{\hspace{0.05cm} z}_{s_{1}mp}(z) = F(a_{1},b_{1},c_{1};z)$ and $F^{\hspace{0.05cm} x}_{s_{2}mp}(x) = F(a_{2},b_{2},c_{2};x)$ depend on $s_{i}$, $m$ and $p$ through the parameters $a_{i}$, $b_{i}$ and $c_{i}$.  These eigenvalues are in agreement with (\ref{laplacian-eigenvalues1}).

\bigskip
\emph{\textbf{Modified eigenvalue problem}}
\bigskip

We shall now look for separable solutions $\Phi^{\pm}(x,z,\xi,\phi)$ to the modified eigenvalue problem, as in (\ref{separable}). Notice that, if
$\Phi^{\mp}(x,z,\xi,\phi)$ is an eigenfunction of $\mathcal{O}_{\mp}$ with eigenvalue $E_{\mp}$, then $\Phi^{\mp}(x,z,-\xi,-\phi)$ is an eigenfunction of $\mathcal{O}_{\pm}$ with the same eigenvalue.  It is therefore sufficient to consider $m$ positive, bearing in mind that the negative $m$ eigenfunctions of $\mathcal{O}_{\pm}$ can be constructed from the positive $m$ eigenfunctions of $\mathcal{O}_{\mp}$.

The resolution of the first eigenvalue problem for $g(z)$ remains unaltered.  The second eigenvalue problem becomes
\begin{eqnarray}
&& \p_{x}\left[x\left(1-x\right)\left(\p_{x}f\right)\right] + \left(1-2x\right)\left(\p_{x}f\right) \\
\nonumber && - \left\{\frac{m\left(m \pm 1\right)\left(1-x\right)}{x} + \frac{q\left(q+1\right)x}{\left(1-x\right)}
+ m\left(m \pm 1\right) + q\left(q+1\right) - E\right\}f = 0.
\end{eqnarray}
We must now distinguish between the $\pm$ signs.  We shall take
\begin{equation}
f(x) = x^{m}\left(1-x\right)^{q} \hspace{0.05cm} h^{+}_{2}(x) \hspace{0.5cm} \textrm{and} \hspace{0.5cm}
f(x) = x^{m-1}\left(1-x\right)^{q} \hspace{0.05cm} h^{-}_{2}(x),
\end{equation}
respectively.  The hypergeometric differential equation for $h_{2}^{+}(x)$ is given by
\begin{eqnarray}
\nonumber && x\left(1-x\right)\p_{x}^{2}h^{+}_{2} + \left[\left(2m+2\right) - \left(2m + 2q + 4\right)x\right]\p_{x}h^{+}_{2} \\
&& - \left\{2\left(mq + m + q\right) + m\left(m+1\right) + q\left(q+1\right) - E \right\}h^{+}_{2} = 0,
\end{eqnarray}
which has solutions $h^{+}_{2}(x) = F(a^{+}_{2},b^{+}_{2},c^{+}_{2};x)$ dependent on the parameters
\begin{equation}
a^{+}_{1}, \hspace{0.05cm} b^{+}_{2} \equiv m + q + \tfrac{3}{2} \pm \sqrt{E + \tfrac{9}{4}}
\hspace{1.0cm} \textrm{and} \hspace{1.0cm}
c^{+}_{1} \equiv  2m + 2. \hspace{1.0cm}
\end{equation}
Here $E = l\left(l+3\right)$, with $l \equiv s_{2} + q + m$ and $s_{2} \in \left\{0, \hspace{0.05cm} 1, \hspace{0.05cm}2, \hspace{0.05cm} \ldots\right\}$, for regularity at $x=1$. Similarly, the hypergeometric differential equation for $h_{2}^{-}(x)$ takes the form
\begin{eqnarray}
\nonumber && x\left(1-x\right)\p_{x}^{2}h^{-}_{2} + \left[2m - \left(2m + 2q + 2\right)x\right]\p_{x}h^{-}_{2} \\
&& - \left\{2\left(mq + m - 1\right) + m\left(m-1\right) + q\left(q+1\right) - E \right\}h^{-}_{2} = 0
\end{eqnarray}
and has solutions $h^{-}_{2}(x) = F(a^{-}_{2},b^{-}_{2},c^{-}_{2};x)$ depending on
\begin{equation}
a^{-}_{1}, \hspace{0.05cm} b^{-}_{2} \equiv m + q + \tfrac{1}{2} \pm \sqrt{E + \tfrac{9}{4}} \hspace{1.0cm} \textrm{and} \hspace{1.0cm}
c^{-}_{1} \equiv  2m, \hspace{1.0cm}
\end{equation}
with $E = l\left(l+3\right) + 2$ and $l$ defined as before.

The eigenfunctions of the modified operator $\mathcal{O}_{\pm}$ can therefore be written as follow:
\begin{equation} \label{dibaryon-mod-eigenfunctions}
\Phi^{\pm}_{smp}\left(x,z,\xi,\phi\right) = z^{\frac{1}{2}|m+p|}\left(1-z\right)^{\frac{1}{2}|m-p|}
x^{m-\frac{1}{2} \pm \frac{1}{2}}\left(1-x\right)^{s_{1} + m} \hspace{-0.05cm} F^{\hspace{0.05cm} z}_{s_{1}mp}(z) \hspace{0.05cm}
F^{\hspace{0.05cm} x \hspace{0.05cm} \pm}_{s_{2}mp}(x)
\hspace{0.05cm} e^{im\xi} \hspace{0.05cm} e^{ip\phi},
\end{equation}
and are associated with the eigenvalues
\begin{equation} \label{dibaryon-mod-eigenvalues}
E^{+}_{smp} = l\left(l+3\right) \hspace{0.5cm} \textrm{and} \hspace{0.75cm} E^{-}_{smp} = l\left(l+3\right)+2,
\hspace{0.5cm} \textrm{with} \hspace{0.5cm} l \equiv s + 2m.
\end{equation}
Here $s = s_{1} + s_{2}$ is a non-negative integer, and $m$ and $p$ are either both integer or both half-integer, with $m \geq |p|$.  Note that $m$ is taken to be non-negative.
The hypergeometric functions $F^{\hspace{0.05cm} z}_{s_{1}mp}(z) = F(a_{1},b_{1},c_{1};z)$ and
$F^{\hspace{0.05cm} x \hspace{0.05cm} \pm}_{s_{2}mp}(z) = F(a^{\pm}_{2},b^{\pm}_{2},c^{\pm}_{2};x)$ were previously described.


\end{document}